\documentclass{acm_proc_article-sp}
\newcommand{\mC}{\mathcal{C}}
\newcommand{\mN}{\mathcal{N}}
\newcommand{\mM}{\mathcal{M}}
\newcommand{\mU}{\mathcal{U}}
\newtheorem{theorem}{Theorem}
\newtheorem{lemma}{Lemma}
\newtheorem{coro}{Corollary}
\newtheorem{prop}{Proposition}
\begin{document}

\title{Node Weighted Scheduling}
\numberofauthors{3}
\author{
\alignauthor Gagan R. Gupta\\
      \affaddr{School of Electrical and Computer Engineering}\\
      \affaddr{Purdue University}\\
     \affaddr{West Lafayette, IN, USA}\\
      \email{grgupta@purdue.edu}
\alignauthor Sujay Sanghavi\\
      \affaddr{School of Electrical and Computer Engineering}\\
      \affaddr{Purdue University}\\
     \affaddr{West Lafayette, IN, USA}\\
      \email{sanghavi@purdue.edu}
\alignauthor Ness B. Shroff\\
       \affaddr{Departments of ECE and CSE}\\
       \affaddr{The Ohio State University}\\
       \affaddr{Columbus, OH, USA}\\
       \email{shroff@ece.osu.edu}
}

\maketitle
\begin{abstract}
This paper proposes a new class of online policies for scheduling in input-buffered crossbar switches. For a system with arrivals, our policies achieve the optimal throughput, with very weak assumptions on the arrival process. For a system without arrivals, our policies drain all packets in the system in the minimal amount of time (providing an online alternative to the batch approach based on Birkhoff-VonNeumann decompositions). Policies in our class are not constrained to be work conserving in every time slot; it may be possible to add edges to the schedule. 

Most algorithms for switch scheduling take an edge based approach; in contrast, we focus on scheduling (a large enough set of) the most congested {\em ports}. This alternate approach allows for lower-complexity algorithms, and also requires a non-standard technique to prove throughput-optimality. One algorithm in our class, Maximum Vertex-weighted Matching (MVM) has worst-case complexity similar to Max-size Matching, and in simulations shows better delay performance than Max-(edge)weighted-Matching (MWM).
\end{abstract}

\section{Introduction}
A commonly used switching fabric in high speed packet switches (e.g., Internet routers) is a crossbar with input queues (IQ) to hold packets during times of congestion. An $N_1\times N_2$ input-buffered crossbar switch contains
$N_1$ input ports and $N_2$ output ports. The crossbar is constrained to schedule a matching i.e., it can send at most one packet from any input port, and receive at most one packet at any output port in a single time slot. The switch scheduling problem is to determine which matching is to be used in every time slot.

\begin{figure}[t]
\centering
\epsfig{file = 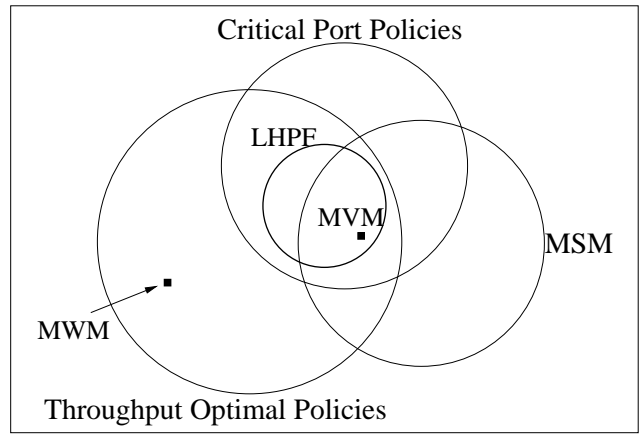, width = 3.3 in}
\caption{\footnotesize Class of scheduling policies for Switches}
\label{fig:policies}
\end{figure}

Most algorithms on switch scheduling take an edge based approach, attempting to schedule either a maximal/maximum set of edges, or those with the largest queues. In this paper we design policies that look only at the weight of the {\em ports} in the switch; queues on the individual edges matter only to the extent that they are non-zero. Intuitively, our policies ensure scheduling of a large enough set of heavy ports in the system. By looking at port weights, we are able to characterize a new class of policies that have lower worst case complexities and are potentially simpler to implement.

To analyze our algorithms, we use a node-based analysis technique. We show that our class of policies is throughput optimal, i.e., they result in stable queues at all admissible loads. We prove throughput optimality using a novel, non-standard Lyapunov function: the maximum total queue at any port. In addition, our policies also achieve {\em minimum clearance time}, i.e., given an initial loading on the switch and no further arrivals, they remove all the packets in the minimum possible time. These policies do not require a priori knowledge of arrival rates.

\subsection{Main Results} \label{sec:main_results}

The focus of this paper is on the design and analysis of policies that determine schedules based upon the total queues at the nodes/ports of the switch. We will construct a class of such policies that are both throughput and clearance-time optimal. Throughout this paper, we will use ``node'' and ``port'' interchangeably. We will also use ``weight'' and ``queue'' interchangeably; they refer to the cumulative queue at the node. For an input port, the queue is the total number of packets waiting to be transferred from the port; for an output port the queue is the number of packets waiting to be transferred to the port. Finally, a matching $M$ is said to match a node $i$ if it contains some edge touching $i$. We now describe the classes of node-weighted policies we investigate.

\begin{figure}[t]
\centering
\epsfig{file = 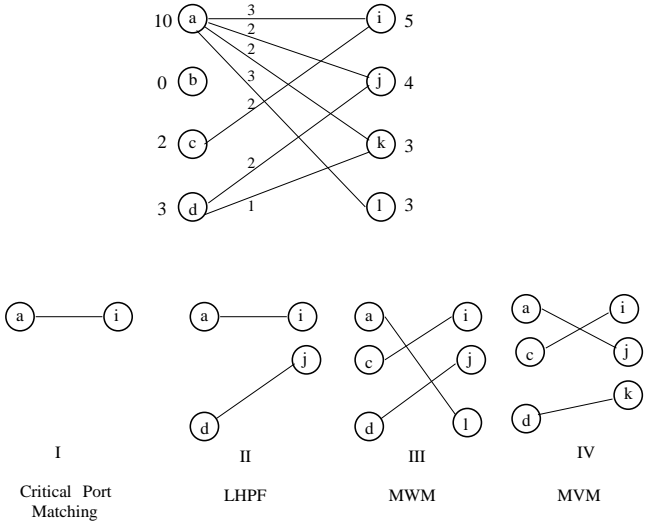, width = 3.35 in}
\caption{\footnotesize Different types of matchings for a switch.}
\label{fig:Ex}
\end{figure}

{\bf Critical Port:} Given a node-weighted bipartite graph, a
port $i$ is {\em critical} if its weight is no smaller than any
other port. A matching $M$ is a {\em critical port matching} if it
matches every critical port. A scheduling policy is a {\em critical
port policy} if it produces a critical matching in every time slot.

{\bf Maximum Vertex-weighted Matching (MVM):} A matching $M$ is an MVM
if the total weight of the nodes it matches is higher than (or equal
to) the total weight of nodes matched by any other matching
$\widehat{M}$. The {\em MVM scheduling policy} is one which schedules
an MVM in every time slot.

{\bf Lazy Heaviest Port First (LHPF):} The {\em threshold} $l(M)$ of a
matching $M$ is the lowest positive integer such that $M$ matches {\em all}
ports with weight greater than or equal to $l(M)$. So, for example,
a perfect matching has threshold 1. $M$ is an LHPF matching if it has the lowest threshold among all possible matchings. We call this the {\em optimal} threshold.  An LHPF {\em policy} is one that produces an LHPF matching in every time slot.

The main result of this paper is that any LHPF policy is throughput-optimal (Thm. \ref{thm:fluid_decrease}). The proof of this result uses a novel Lyapunov function: the weight of the heaviest port. We also show that a policy is clearance time optimal iff it is a critical port policy (Prop.~\ref{prop:clearance_critical}). Given any queue configuration, a critical port matching always exists; we also provide a simple way to find it. This enables us to develop a ``slot-by-slot'' algorithm for the clearance problem, as opposed to existing batch policies \cite{Mc02, Neely07, weller} based on Birkhoff-VonNeumann decompositions.

We call our class ``lazy'' because a LHPF matching may not even be maximal; in particular, it may not match any extra nodes below the optimal threshold (beyond what it needs to satisfy those above the threshold). To clarify our classes of policies we give a simple example in Fig.~\ref{fig:Ex}.

Consider a 4$\times$4 IQ switch with edge weights and corresponding port weights as shown in the Fig.~\ref{fig:Ex}. There is only one critical port (port {\em a}) with weight 10. So a critical port policy must at least schedule port {\em a}. Now let us consider a LHPF matching. It is clear that the size of a matching can at most be three. It follows that not all the ports on the output side can be matched, in particular the threshold must be strictly greater than 3. Hence any matching that at least schedules ports {\em a}, {\em i} and {\em j} is LHPF. For example II, III and IV are LHPF matchings. There is a unique MWM (III) in the graph. There are many MVMs in this graph. III and IV are both MVM.

Fig.~\ref{fig:policies} shows how the different policy classes relate to each other. It is well known that MWM \cite{ DaiPra, Sasha04, shah06, tassiulas92, Tass} and MVM \cite{Mc98} are throughput optimal. It is also known that the MVM policy is a maximum-size matching policy \cite{Mc98}, but that not all MSM policies are throughput optimal \cite{Mc96, Mc03}. For the policy classes defined in this paper, critical policies need not be throughput optimal. Lemma~\ref{lem:lhpf_is_critical} shows that any LHPF policy is also a critical port policy. Theorem~\ref{thm:fluid_decrease} shows that any LHPF policy is throughput optimal. Corollary~\ref{coro:mvm_is_lhpf} shows that any MVM is an LHPF matching. In Section~\ref{sec:clear_time} we provide an example to show that the clearance time of popular existing policies (like MWM, MSM and Greedy weighted maximal matching(GMM)) may be as large as twice the optimal. 

We now discuss some implications of our work from an algorithmic perspective. We prove simple properties of LHPF policies which can be used as a source for algorithms to find LHPF matchings in node-weighted graphs. This is similar to augmenting-path characterizations which provide algorithms for edge-weighted matching problems (like maximum cardinality, maximum-edge-weight etc.). We provide a way to modify simple but non-throughput-optimal policies (like edge-based greedy, or maximal matching) into throughput optimal ones via post-processing. We elaborate on this procedure in Section~\ref{sec:policies}.

The tradeoff between delay and implementation complexity has been studied in \cite{Neely02c, Neely07, Mc98}. In general, a lower complexity scheduler will result in higher delays. There are simple algorithms like maximal matching and GMM, which empirically perform well in most cases \cite{Joo07, Joo08, shah03}, but are difficult to analyze. In fact they are not even throughput optimal in some cases. When used in conjunction with LHPF (via post-processing), they should have both good delay and throughput. Note that one of the members of LHPF class is the MVM algorithm which can be shown empirically to have delay performance very close to the well known delay-optimal MWM-$\alpha$ \cite{Mc01, Sasha04, shah06} policies at a lower complexity \cite{Mc98} of $O(N^{2.5})$ as compared to $O(N^{3})$ for MWM \cite{karp}. LHPF class contains policies which are simpler to implement than the MVM and hence are potential candidate for a low complexity delay efficient scheduler with theoretical guarantees on the throughput. Additionally, these policies are clearance-time optimal.

\subsection{Related Work} \label{sec:related}
Throughput optimal policies can be classified broadly into Backlog-aware policies, which require the knowledge of the backlogs at every time slot and those which are Backlog independent.

Backlog independent policies instead use the knowledge of the arrival rates \cite{Altman, Chang} to construct a randomized or periodic scheduling rule precisely matched for the input rates. Such scheduling offers arbitrarily low per-time slot computation complexity at the cost of large delays (shown to be O(N) \cite{Neely07}, where N is the size of switch).

Backlog aware policies can be further classified into those which are {\em frame} based or {\em batch} based and the {\em online} policies.

The {\em frame} based policies are considered in \cite{weller, Mc02, Neely07} and are based on the principle of iteratively clearing the backlog in minimum time. The throughput optimality of these policies is restricted to Bernoulli i.i.d. traffic in \cite{weller, Mc02}. In \cite{Neely07}, prior knowledge of the statistics of arrival process is required to be able to select the {\em frame} size appropriately so as to achieve throughput optimality. Minimum clearance time policies have been applied to stabilize networks in \cite{Neely02a, Gamarnik}. {\em Batch} based policies \cite{Mc02} are similar to the {\em frame} based policies except that the {\em frame} size is dependent on the traffic arrival pattern and the scheduling algorithm used.

In this paper we restrict our attention to the development of {\em online} algorithms, which attempt to schedule traffic by computing a matching every time slot. One such policy is the famous MWM policy which computes the maximum weight matching and is known to be throughput optimal. The proof for stability can be provided either in the fluid limits \cite{DaiPra} or in the stochastic sense\cite{Tass}. But essentially it hinges on a quadratic Lyapunov function and ensuring that the drift is negative.

The Maximum Size Matching (MSM) policy schedules the maximum size matching and hence maximizes the instantaneous throughput in each time slot. However it is known that if ties are broken randomly, MSM does not achieve 100\% throughput for all admissible Bernoulli traffic patterns \cite{Mc96, Mc03}. It is possible that if the ties are broken carefully, a special MSM might be stable. Among the class of MSM policies, there are two polices that have been proposed in the literature to be throughput optimal: MVM  and MWM-0+. \\
{\bf MVM} is known to be throughput optimal \cite{Mc98}. The proof of throughput optimality in \cite{Mc98} uses the fact that a MVM on a graph $G$ is a MWM on a graph $G^{'}$, where edge weights have been selected carefully.
The technique to prove throughput optimality of MVM is essentially the same as that for MWM. The proof provided in this paper serves as a alternate, since MVM is a member of the LHPF class of policies. \\
{\bf MWM-0+ :} At each time slot, consider all matchings which have maximal size. Among these choose one which has maximum weight, with weight function log. Break ties arbitrarily. This is conjectured to be throughput optimal in \cite{shah06}.\\

It is useful to also consider online scheduling according to maximal matches, which are matchings where no new edges can be added without sharing a node with an already matched edge. Maximal matchings can be found with O($N^{2}$) operations and the computation is easily parallelizable to O(N) complexity \cite{weller}. Greedy weighted maximal matching (GMM) is a scheduler that tries to schedule the heavy edges. The GMM policy has been analyzed for the general class of networks with interference constraints \cite{dimakis06} where it is shown that they achieve full throughput in a network that satisfies the local pooling condition. In simple terms, the local pooling condition means that a vector $\lambda$ in the capacity region cannot dominate another vector $\mu$ in the capacity region in all the coordinates. This result can be generalized \cite{Joo07, Joo08, Bree08} to show that GMM achieves at least a certain fraction of the capacity region given by the local pooling factor. Although our Lyapunov function looks similar to that in \cite{dimakis06, Joo07, Joo08, Bree08} it is based on node weights as opposed to weights on the individual edges in the graph. Moreover, we can show that the LHPF class of policies are not even required to be maximal in every time-slot whereas the policies considered in \cite{Joo07, Joo08, Bree08} are.
%required to be maximal in every time-slot.

The general research on the delay analysis of scheduling policies has progressed in the following main directions:
\begin {itemize}
\item{ {\it Heavy traffic regime using fluid models:} Fluid models have typically been used to either establish stability of the system or to study the workload process in the heavy traffic regime. It has been shown in \cite{Sasha04} that the MWM policy minimizes the workload process for a generalized switch. Furthermore, \cite{shah06} proves multiplicative state space collapse of a family of scheduling algorithms related to MWM in the heavy traffic regime and conjectures an optimal algorithm MWM-0+.
}
\item{ {\it Stochastic Bounds using Lyapunov drifts:}  This method is developed in \cite{ leonardi01, NOW, Neely02c} and is used to derive upper bounds on the average queue length for these systems. However, these results are order results and provide only a limited characterization of the delay of the system. For example, it has been shown in \cite{Neely07} that the bounds in \cite{leonardi01, shah02} are O(N) bounds and hence not very useful. It is also shown that it is possible to achieve O(log N) delay.}
\end{itemize}

As noted in \cite{Mc98}, the MVM policy combines the benefit a maximum size algorithm, with those of a maximum weight algorithm, while lending itself to simple implementation in hardware. In MVM, each weight is a function of queue lengths (sum of all edges that touch a node) and hence it has an advantage of both the maximum size matchings with high instantaneous throughput while guaranteeing high throughput, even when the arrival traffic is non-uniform. We have in fact characterized a class of policies much larger than the MVM policy and potentially lower complexity and equivalent performance benefits.

\section{Preliminaries} \label{sec:prelim}

\noindent {\bf Switches:} This paper is about scheduling in (the
standard) input-buffered crossbar switches, which we now briefly
describe. An $N_1\times N_2$ input-buffered crossbar switch contains
$N_1$ input ports and $N_2$ output ports. The system operates in
discrete time slots. In each slot, packets may arrive at the input
ports; each packet has an output port it needs to be transferred
to. Packets have to be transferred from inputs to outputs, under the
following constraint: in any one time slot each input port can send at
most one packet to at most one output port, and each output port can
receive at most one packet from at most one input port. The scheduling
problem is to determine how to transfer packets subject to these
constraints.

{\bf Notation:} Switch scheduling can be modeled as the problem of
finding matchings in bipartite graphs, one in every time slot. Consider $G(s)$ the graph at slot s. $G(s)$ is a
bipartite graph with input ports on one side and output ports on the other. As mentioned in the introduction, we will use
``nodes'' and ``ports'' interchangeably. There is an edge $(i,j)$ in $G(s)$
if and only if there is at least one packet at input $i$ that has output
$j$ as its destination. The scheduling algorithm
finds a matching $M(s)$ in $G(s)$; then, for every edge $(i,j)\in
M(s)$ one packet is then transferred from $i$ to $j$. These packets
are then considered to have left the system. A scheduling policy is a
rule to pick the matching $M(s)$, in every slot $s$, based on the state of
the system. For any input port $i$, $q_i(s)$ denotes the {\em total}
number of packets at $i$. Similarly, for any output port $j$, $q_j(s)$
denotes the total number of packets in the system (i.e. all inputs)
that are waiting to be transferred to $j$. We will not need to refer
to the queues on individual edges. We will however often refer to the
total queue at a port as the ``weight'' of that port; ``heavy'' ports
have more packets in their queues than ``lighter'' ports. 
%We now define the class of policies we investigate in this paper.

%We show in Proposition \ref{prop:mvm_lhpf} that any MVM is an LHPF matching, and thus MVM scheduling is an LHPF policy.

We now state a couple of well-known results, from \cite{Hall, Perfect, Kuhn} which we will use in the
proofs of this paper.

\begin{lemma}[Hall's Condition]\label{lem:halls}
Let $G$ be any bipartite graph, with the two partitions being $V_1$
and $V_2$. Let $S_1\subset V_1$ be any subset of one partition.  Then,
there exists a matching in $G$ that matches every node in $S_1$ if and
only if for every further subset $S\subset S_1$, we have that
$|\mN(S)| \geq |S|$. Here the neighborhood $\mN(S)$ is all nodes in
$V_2$ that have an edge to some node in $S$.
\end{lemma}

\begin{lemma}\label{lem:merge}
Let $G$ be any bipartite graph, with the two partitions being $V_1$
and $V_2$. Let $S_1\subset V_1$, and suppose there exists a matching
$M_1$ that matches all nodes in $S_1$. Similarly, let $S_2\subset V_2$
and there exist and $M_2$ that matches all nodes in $S_2$. Then there
exists a matching $M$ that matches all nodes in {\em both} $S_1$ and
$S_2$.
\end{lemma}

Note that in Lemma \ref{lem:merge}, $M$ may not match the nodes in the
two sets to each other; just that each node in $S_1\cup S_2$ will be
matched to some node in the graph.

{\bf Graph-theoretic preliminaries:}\\
We now formally define the terms
we will use. All are standard, except for the definition of
``absorbing paths''. Throughout, we consider a node-weighted
graph. The {\em length} of a path is the number of edges it
contains. The {\em weight $w(M)$ of a matching $M$} is the total
weight of all the nodes it matches. For any two matchings $M_1$ and
$M_2$, the {\em symmetric difference}, denoted by $M_1\triangle M_2$, is
the set of edges in one of the two matchings, but not in both. It
is well known that $M_1\triangle M_2$ is always the node-disjoint
union of paths and even-length cycles. Finally, given a matching $M$
and path $P$, the set $M\oplus P = M - (M\cap P) + (M^c \cap P)$
denotes the edges obtained by ``flipping'' the edges in $P$. We now
define the two scenarios of our interest where the resulting set
$M\oplus P$ is also a matching.

Given a matching $M$, and any node $i$ not matched by $M$,
\begin{enumerate}
\item An {\em augmenting path from $i$} is any odd-length path $P$
  whose every alternate edge is in $M$, has $i$ as one endpoint, and
  ends at an unmatched node (say $j$).

  Note that now $M\oplus P$ matches every node $M$ does, and in
  addition matches $i$ and $j$ as well. Thus its weight is $w(M\oplus
  P) = w(M) + w_i + w_j$, which is strictly bigger than $w(M)$.
\item An {\em absorbing path from $i$} is any even-length path $P$
  whose every alternate edge is in $M$, has $i$ as one endpoint, and
  whose last endpoint -- say $j$ -- has weight $w_j < w_i$.

  Note that now $M\oplus P$ matches every node $M$ does except $j$,
  which is replaced by $i$.  Thus it has strictly higher weight:
  $w(M\oplus P) = w(M) + w_i - w_j > w(M)$
\end{enumerate}

\begin{figure}[t]
\centering
\epsfig{file = 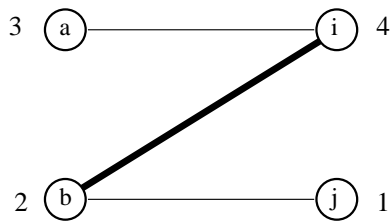, width = 2 in}
\caption{\footnotesize Augmenting and absorbing paths. Edge $(b, i)$ is in the matching M. $a-i-b$ is an absorbing path. $a-i-b-j$ is an augmenting path.}
\label{fig:aug}
\end{figure}

Fig.~\ref{fig:aug} illustrates the idea of augmenting and absorbing paths. $a-i-b$ is an absorbing path from $a$ since it is an even-length path ending in a node with smaller weight. $a-i-b-j$ is an augmenting path from $a$ since it is a odd-length path and ends in an unmatched node $j$. 

\section{Clearance Time and Critical Port Policies}\label{sec:clearance}

In the clearance time problem, the queues in the system have an
initial loading, and there are {\em no arrivals}. We are interested in
scheduling so as to minimize the {\em clearance time}, which is the
time before every packet in the initial loading has exited the
system. In the following, $q_i(s)$ denotes the remaining packets at
port $i$ immediately after time slot $s$, and $q_i(0)$ the initial
loading.

Since at most one packet can be scheduled at any given port, an
obvious lower bound on the clearance time is
\begin{equation}
\label{eq:clearance_lower}
\tau ~ \geq ~ \max_i q_i(0)
\end{equation}
It is known that this lower bound is tight, based on the following
``batch'' policy. We first briefly describe this policy, and then describe a
more elegant slot-by-slot policy. Let $\tau^* = \max_i q_i(0)$.

{\bf Batch policy:} This is based on the the Birkhoff-VonNeumann
theorem. Let $N=\max\{N_1,N_2\}$, and consider the $N\times N$ matrix
$L$ in which, for $i\leq N_1$ and $j\leq N_2$, has entries $L(i,j) =
\frac{q_{ij}(0)}{\tau^*}$, and $q_{ij}(0)$ is the number of packets
waiting at input $i$ for output $j$ in the initial loading. All the
other entries of $L$, i.e. all $L(i,j)$ for which either $i> N_1$ or
$j> N_2$, are 0. It is clear that $L$ is a sub-stochastic matrix
(i.e., the sum of every row and every column is less than or equal to
1). The Birkhoff-VonNeumann theorem says that any such matrix can be
represented as a convex combination of (sub)permutation matrices; each
(sub)permutation matrix corresponds to a matching in the
switch. Furthermore, the fact that every entry of $L$ is an integer
multiple of $\frac{1}{\tau^*}$ implies that a batch of at most $q^*$
such matchings will be needed.  Thus the lower bound is tight, and can
be achieved by this batch of matchings \cite{weller, Mc02, Neely07}. 

The Birkhoff-VonNeumann approach above gives us an algorithm for
clearing out a given batch of packets, but it would be more practical to
have a ``slot-by-slot'' solution: one in which the matching at each
time can be easily determined from the current loading.  We now show
that the class of critical port policies is exactly what is needed for
a slot-by-slot solution.

\begin{prop}\label{prop:clearance_critical}
A scheduling policy is clearance-time optimal, i.e. it achieves the
lower bound (\ref{eq:clearance_lower}), if and only if it is a
critical port policy.
\end{prop}

{\em Proof:} Suppose $\pi$ is a clearance-time optimal policy. This
means that at any time slot $s<\tau^*$, every port $i$ has $q_i(s) \leq
\tau^* -s$; otherwise, the port cannot be emptied by time slot $\tau^*$.
Also, it is clear that all the ports with initial
load $q_i(0) = \tau^*$ will now have $q_i(s) = \tau^* -s$; thus the
weight of the critical ports at time slot $s$ is $\tau^* -s$. If any
one of these critical ports is excluded by $\pi$ in slot $s$, it will
have a total queue of $\tau^*-s$ at time slot $s+1$, and hence cannot
be drained by time $\tau^*$. Thus every clearance-time optimal policy
is a critical port policy.

Conversely, suppose now that $\pi$ is a critical-port policy. It is
easy to see that in any time slot the maximum load at any port will
decrease by exactly one. This is because the ports with the maximum
loads are the critical ports, and every one of them will be scheduled
by $\pi$ in slot $s$. \hfill $\blacksquare$

\begin{coro}
\label{coro:cpm}
Given any set of queues, there exists a critical-port matching.
\end{coro}

{\em Proof:} Given the set of queues, consider the clearance time
problem with these queues as the initial loading. We know that there
exists a policy, e.g., based on the Birkhoff-VonNeumann decomposition,
that achieves the bound (\ref{eq:clearance_lower}). By Lemma
\ref{prop:clearance_critical}, this policy has to be a critical-port
policy. Hence, in the first time slot it will have a critical-port
matching. This implies such a matching exists for our set of queues.
\hfill $\blacksquare$

We now give a procedure to find a critical-port matching, given any
set of queues. 
%The procedure serves as an alternative constructive proof to the above Corollary.

Procedure for Critical Port Matching
\hrule
INPUT: A node-weighted graph, and any initial matching $M_0$ (which could be empty) \\
OUTPUT: $M^*$, a critical-port matching \\
\begin{itemize}
\item Set $l$=1
\item While there exists critical port $i$ not matched by $M_{l-1}$, 
    \begin{itemize}
    \item Find $P$, an augmenting path or absorbing path from $i$ with respect to $M_{l-1}$.
    \item Set $M_l = M_{l-1}\oplus P$ and increment $l = l+1$
    \end{itemize}
\end{itemize}
\hrule

\begin{lemma}\label{lem:crit_aug_abs}
Given any matching $M$, and a critical port $i$ not matched by $M$, there exists an augmenting path or alternating path $P$ from $i$. 
\end{lemma}

{\em Proof:} By Corollary \ref{coro:cpm}, there exists a matching $M^*$ that matches all critical ports. In particular, it matches $i$. Consider now the symmetric difference $M\triangle M^*$, which contains node-disjoint paths and cycles; since $i$ is not matched by $M$, $i$ will be the endpoint of a path $P$ in $M\triangle M^*$. If $P$ is of odd length, it is an augmenting path, and we are done. If $P$ is even length, let $j$ be the other endpoint of $P$. Now, $P$ begins at $i$ with an edge in $M^*$, so it ends in $j$ with an edge in $M$. Also, there is no edge in $M^*$ touching $j$, because $j$ is the endpoint in the symmetric difference. This means that $j$ cannot be a critical port, because $M^*$ matches every critical port. Since $i$ is critical, this means that $w_i > w_j$, which means that $P$ is an absorbing path. \hfill $\blacksquare$

{\bf Correctness of Procedure:} Suppose at iteration $l$, we have that $M_{l-1}$ does not match critical port $i$. Lemma \ref{lem:crit_aug_abs} guarantees that an augmenting or absorbing path $P$ from $i$ will be found. Also, if $M_l = P\oplus M_{l-1}$ then $i$ will be matched by $M_l$. Thus all we need to show is that any critical port that is matched by $M_{l-1}$ remains matched by $M_l$. This is so because: if $P$ is augmenting, $M_l$ matches all nodes matched by $M_{l-1}$. If $P$ is absorbing, the node $j$ removed at the expense of $i$ is not critical, because absorbing requires that $w_j < w_i$. Thus the procedure gives us the desired critical port matching.

\subsection{Clearance-time of other Policies} \label{sec:clear_time}
\begin{figure}[t]
\centering
\epsfig{file = 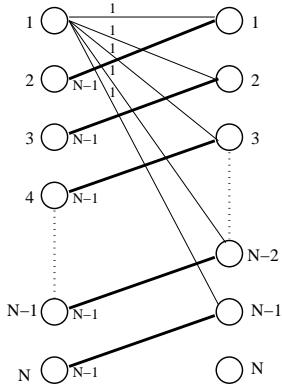, height = 2 in}
\caption{\footnotesize An example where MWM, GMM, MSM and MWM-0+ are not clearance-time optimal}
\label{fig:Main}
\end{figure}

We now provide an example to show that edge weight based policies like MWM, Greedy weighted maximum matching (GMM) and MSM are not clearance time optimal.

Consider a N $\times$ N switch with the following configuration. Input Port $1$ has one packet each destined for ports $1$ through $N-1$. Port $i,~ 2\geq i \geq N$ have $N-1$ packets destined for output port $i-1$. The clearance time $\tau^{*}$ for the above configuration is $N$.

Let us consider, how MWM schedules packets in the given system. In the given system, at any time, no more than $N-1$ input ports can be matched under the switch constraints. The maximum weight matching policy does not match input port $1$ for the first $N-2$ slots since for any output port $j$, the edge weight $q_{1j}$ is smaller than $q_{(j+1)j}$.
So, after $N-2$ slots, the weight of input port $1$ is $N-1$. Depending on how ties are broken, MWM will take either $N$ or $N-1$ more slots to clear all the packets in the system. Hence, MWM clears the packets in at least $2N-3$ slots whereas the clearance time is $N$. This example shows that MWM can take twice as much time to clear the system for large $N$. The GMM, MWM-$\alpha$ and MWM-$0+$ policies will schedule this system in the same manner as discussed above and take at least $2N-3$ slots to empty all the packets in the system. 

What is attractive is that LHPF policies being critical port policies are also clearance-time optimal, which we prove in the next lemma.

\begin{lemma}
\label{lem:lhpf_is_critical}
Any LHPF matching is also a Critical port matching, and hence any LHPF
policy is also a Critical port policy.
\end{lemma}

{\em Proof:} By Corollary~\ref{coro:cpm}, for any set of queues, there exists a critical port matching.
Hence the optimal threshold (defined in Section~\ref{sec:main_results} ) must be smaller than the $\tau^{*}$, the weight of the critical port. Since the LHPF policy matches all ports above the optimal threshold, it will
match all critical ports and hence is a Critical port policy.
\hfill $\blacksquare$

\section{LHPF Policies} \label{sec:policies}

We now take a closer look at LHPF matchings, and LHPF policies. Recall the definition of threshold l(.) from the introduction. LHPF matchings were defined in the introduction as those in which all the ports whose weight is above the optimal threshold are matched. The optimal threshold is the lowest possible threshold for a graph.
We now present a structural {\em sufficient} condition for a matching
to be LHPF. This will enable us to both develop algorithms for LHPF
matchings, and to understand their properties.

\begin{lemma}\label{lem:LHPF}
  $M$ is an LHPF if at least one of its {\em heaviest} unmatched nodes
  has no augmenting path or absorbing path.
\end{lemma}

{\bf Remarks:} Note that the condition just concerns one of the
heaviest unmatched nodes; every other unmatched node (heaviest or
otherwise) is free to have augmenting/alternating paths. This is a
reflection of the fact that LHPF matchings need not even be maximal.

Lemma \ref{lem:LHPF} is a sufficient condition for a matching to be
LHPF, but it is not necessary. This is because if the {\em heaviest} unmatched node is below the optimal threshold, then it is not required to be matched to be LHPF. For example, consider the graph in Fig.~\ref{fig:Ex}. Matching II, for example is a LHPF, although there is an augmenting path $l-a-i-c$ which results in matching III, which is again LHPF.

{\em Proof:} We will prove the contrapositive, i.e. we will prove that
if $M$ is {\em not} and LHPF then {\em every} heaviest unmatched node
will have an augmenting or absorbing path.  Let $w$ be the weight of
the heaviest node not matched by $M$, and let $\mU$ be the set of
heaviest unmatched nodes (i.e., all unmatched nodes with weight $w$).

Now, by assumption, $M$ is not LHPF. Let $M^*$ be any LHPF
matching. It follows that the threshold of $M^*$ is strictly lower
than that of $M$, which can only happen if $M^*$ schedules all nodes
of weight $w$, and in particular, all nodes in the set $\mU$. Consider
now the symmetric difference $M\triangle M^*$, which contains
node-disjoint paths and cycles. Every $i\in \mU$ is matched by $M^*$
but not by $M$. Thus each $i\in \mU$ will be an {\em endpoint} of a
path, say $P_i$, in $M\triangle M^*$.

Consider any such $i$ and $P_i$.  If $P_i$ is of odd length, it is an
augmenting path. If $P_i$ is of even length, let $j$ be its other
endpoint. Because $P$ is even length, $j$ is not matched by
$M^*$. This means that $w_j < w = w_i$, which implies that $P_i$ is an
absorbing path. \hfill $\blacksquare$

In the same way that augmenting-path characterizations provide
algorithms for edge-weighted matching problems (like maximum
cardinality, maximum-edge-weight etc.), Lemma \ref{lem:LHPF} can be
used as a source for algorithms to find LHPF matchings in
node-weighted graphs; it can also be used to modify a (potentially
non-LHPF) matching to obtain an LHPF one. We now describe a simple
procedure for either of these tasks.

Procedure for LHPF
\hrule
INPUT: a node-weighted graph, and any initial matching $M_0$
(which could be empty, or generated by some other algorithm) \\
OUTPUT: $M^*$, an LHPF matching
\begin{itemize}
\item Set $l = 1$
\item At iteration $l$,
  \begin{itemize}
  \item IF $M_{l-1}$ matches all nodes, set $M^* = M_{l-1}$ and BREAK.
  \item Pick any highest unmatched node $i$ in $M_{l-1}$, and try to
    find an augmenting or absorbing path $P$ from $i$.
  \item IF such a $P$ can be found, set $M_l=M_{l-1}\oplus P$ and
    $l=l+1$.
  \item ELSE set $M^* = M_{l-1}$, and BREAK loop.
  \end{itemize}
\end{itemize}
\hrule

{\bf Remarks:} The above description is just a conceptual procedure;
efficient implementations could potentially rely on optimizations
(e.g. like parallelism, as was done in \cite{karp} for
max-cardinality matching). We emphasize, rather, a more interesting
aspect of the above procedure: is that it allows us to make LHPF
matchings out of non-LHPF ones via post-processing. In particular, given a matching $M_{0}$, by a simple policy, one can go down the sequence of ports, and add them to the current matching: either via augmentations, or at the expense of some node with strictly lower weight. The procedure stops at the point of first failure to find $P$; it can be expected, on average, that if the initial matching is maximal (say for example if it is the greedy matching), then the number of nodes $v$ that need to be inspected may be small. As mentioned in Section~\ref{sec:main_results}, this could be of advantage in several settings.

This procedure can also be used as a pre-processing step beginning with an empty matching to give a LHPF matching with possibly many unmatched nodes (those which are below the optimal threshold). This matching can be extended by a low complexity scheduler, for example to a maximal matching to improve the delay performance.

{\bf Correctness of Procedure:} The correctness of the above
sequential adding procedure follows from Lemma \ref{lem:LHPF}. In
particular, if there are no unmatched nodes, then the matching is
clearly LHPF. If the procedure stops at iteration $i$, it means that
for the matching $M_{i-1}$, and a heaviest unmatched node $v$, there
is no augmenting or absorbing path from $v$. This is exactly the
condition under which Lemma \ref{lem:LHPF} guarantees that the
matching is LHPF.

For clarity, we now describe a policy that is {\em not} an LHPF
policy. Suppose we do the following: go down the sequence of ports,
recursively matching nodes if any neighbor is free, but not changing
the edges already previously matched. Even if we go all the way to the
end, this policy is not LHPF because it may exclude a port that would
have been possible to schedule by changing the matchings of heavier
ports that came before it. It is thus important that the ports are
added via augmenting or absorbing paths.

One example of an LHPF matching that has been previously studied is
Maximum Vertex-weighted Matching.

\begin{lemma}\label{lem:MVM}
$M$ is an MVM if and only if there is no augmenting path or
  absorbing path from any of its unmatched nodes.
\end{lemma}

{\bf Remarks:} Contrast Lemma \ref{lem:MVM} to Lemma \ref{lem:LHPF}.
MVM requires that {\em all} unmatched nodes have no augmenting or
absorbing path. LHPF requires only that one of the heaviest unmatched
ports satisfy this property.

{\em Proof:} Let $M$ be a MVM. If there exists some path $P$ that is
either an augmenting path or an absorbing path, then the matching
$M\oplus P$ will have strictly higher weight then $M$, which is a
contradiction. Thus no such $P$ exists.

Now suppose that $M$ has no augmenting or alternating paths. Suppose
also that it is not a MVM. Let $M^*$ be any MVM, and consider
$M\triangle M^*$; it is a collection of node-disjoint cycles and
paths. Consider any path $P$ in this collection. If $P$ is of odd
length, it is either an augmenting path for $M^*$, or for $M$. The
latter possibility is ruled out by assumption, and we just proved that
the former is ruled out too: $M^*$ is an MVM, and so cannot have an
augmenting path. So there are no odd length paths. This means there
has to be an even length path $P_1$ in the collection whose endpoints
have unequal weights; else the weights of $M$ and $M^*$ will be
equal. However, depending on which endpoint is heavier, this $P_1$ is
either an absorbing path for $M$ or $M^*$; again, either possibility
is ruled out. \hfill $\blacksquare$

\begin{coro}\label{coro:mvm_is_lhpf}
Any MVM matching is an LHPF matching, and hence the MVM scheduling
policy is an LHPF policy.
\end{coro}

{\em Proof:} By Lemma \ref{lem:MVM}, any MVM will not have an
augmenting or absorbing path from any unmatched node; including any of
its heaviest unmatched nodes. By Lemma \ref{lem:LHPF} this implies
that it is an LHPF matching. \hfill $\blacksquare$

The complexity of MVM is $O(N^{2.5})$ \cite{Mc98} and the policy is simple to implement in hardware. Many heuristics have been developed for MSM and they can be readily tuned to compute approximate MVMs. with the characterization of LHPF policies, which is a much bigger class, we expect that it would be much easier to develop heuristics for LHPF matchings.
MSM is a special case of MWM. The proof uses the fact that a MVM is a MWM on a graph where edge weight on an edge connecting input node $i$ and output node $j$ have been chosen as follows:\\
\begin{equation}
 w_{ij}(n) =  \left \{ \begin{array}{ll}
  \mbox{$q_{i}(n)+q_{j}(n)$,} & \mbox{ if $q_{ij}(n)>0$}\\
  0, & \mbox{ otherwise }
\end{array}
\right.
\label{sched}
\end{equation}

\begin{lemma}
If $G$ has a perfect matching (i.e. one that matches every port), then
any LHPF matching also has to be perfect.
\end{lemma}

{\em Proof:} The existence of a perfect matching means that the
optimal threshold for a matching is 1. Any non-perfect matching will
have a higher threshold, and hence not be an LHPF. \hfill $\blacksquare$

\section{Throughput Optimality of LHPF policies}

In this section we show that any LHPF policy is throughput optimal. Let the system be empty at time 0. Let $a_{i}(n)$ denote the cumulative number of packets that have arrived at an input port $i$ up to time slot $n$. Similarly, $a_{j}(n)$ denotes the cumulative number of packets that have arrived in the system, destined for output port $j$ up to time slot $n$. For each edge in the matching, one packet is removed at both the nodes touching the edge. With this understanding, henceforth, we shall not distinguish between an output and input port. We assume the convention that $a_{i}(0)=0$. We assume that the arrival processes $a_{i}(.)$ satisfy a strong law of large numbers (SLLN): with probability one,
\begin{equation}
\label{fluidA}
\lim_{n \rightarrow \infty} \frac{a_{i}(n)}{n} = \lambda_{i}
\end{equation}

For any port, input or output, let $\lambda_i$ be the average rate of arrival of packets to port $i$. Define
\[
\epsilon^* ~ = ~ \min_i \, (1-\lambda_i)
\]
The capacity region is $\{\mathbf{\lambda}: \lambda_i <1 \,\, \text{for all
  $i$} \}$, which means that $\epsilon^* >0$.

{\bf Fluid Model}

We develop a fluid limit model following the development in \cite{DaiPra}.
%Let $a_{i}(n)$ denote the cumulative number of packets that have arrived at an input port $i$ up to time slot n.
Let $q_{i}(n)$ denote the weight at port $i$ and $d_{i}(n)$ be the number of packets that departed from port $i$ by time slot $n$. Let $h_{M}(n)$ be the number of slots in which matching $M \in \mM$ has been scheduled, where $\mM$ is the set of all matchings (not necessarily maximal). Then $h_{M}$ is a non-decreasing function. Also note that by definition of $G(n)$, $M$ can schedule only non-zero edges in the system. $M_{i}$ indicates if matching $M$ schedules port $i$. Note that $q_{i}(.)$ and $d_{i}(.)$ evolve according to the following:

\begin{equation*}
q_{i}(n)= q_{i}(0) + a_{i}(n)-d_{i}(n)\\
\end{equation*}
\begin{equation*}
d_{i}(n)= \sum_{M \in \mM} \sum_{l=1}^{n} M_{i}(h_{M}(l) - h_{M}(l-1))\\
\end{equation*}
\begin{equation*}
\sum_{M \in \mM} h_{M}(n) = n
\end{equation*}

We define $a_{i}(t)$ for a non-negative real number $t$ by interpolating the value of $a_{i}$ between time $\lfloor t \rfloor$ and $\lfloor t \rfloor +1$. We also define $q_{i}(t)$ and $d_{i}(t)$ in the same way by linear interpolation of the corresponding values at time $\lfloor t \rfloor$ and $\lfloor t \rfloor + 1$. Then, by using the techniques of Theorem 4.1 of \cite{Dai}, we can show that, for almost all sample paths and for all positive sequence $x_{k} \rightarrow \infty$, there exists a subsequence $x_{k_{l}}$ with  $x_{k_{l}} \rightarrow \infty$ such that the following convergence holds uniformly over compact intervals of time $t$:

For all $i$,
\begin{eqnarray}
\frac{a_{i}(x_{k_{l}}t)}{x_{k_{l}}t}\rightarrow \lambda_{i} t & \quad \quad & \frac {q_{i}(x_{k_{l}}t)}{ x_{k_{l}}} \rightarrow Q_{i}(t) \\
\frac {d_{i}(x_{k_{l}}t)}{ x_{k_{l}}}   \rightarrow D_{i}(t) &&
\frac {h_{i}(x_{k_{l}}t)}{ x_{k_{l}}}   \rightarrow H_{i}(t) \\
\end{eqnarray}

The system $(D, H, Q)$ is called the fluid limit and queues evolve in the fluid limit as follows:
\begin{equation*}
Q_{i}(t) = Q_{i}(0) + \lambda_{i}(t) - D_{i}(t)\\
\end{equation*}
\begin{equation*}
\frac{d}{dt} D_{i}(t) = \sum_{M \in \mM} M_{i} \frac{d}{dt} H_{M}(t)\\
\end{equation*}
\begin{equation*}
\sum_{M \in \mM} H_{M}(t) = t
\end{equation*}
$D, H $ and $Q$ are absolutely continuous functions and are differentiable at almost all times $t \geq 0$ (called {\em regular} times). It follows that
\begin{equation}
\label{eq:der}
\begin{split}
\frac{d}{dt} Q_{i}(t) &= \lambda_{i} - \frac{d}{dt}  D_{i}(t)\\
                                            &= \lambda_{i} - \sum_{M \in \mM} M_{i} \frac{d}{dt} H_{M}(t)\\
\end{split}
\end{equation}

The following lemma from \cite{DaiPra} establishes the connection between the stability of the switch and the fluid model.
\begin{lemma}
A switch operating under a matching algorithm is rate stable if the corresponding fluid model is weakly stable.
\end{lemma}

\begin{lemma}
The fluid model of a switch operating under a matching algorithm is weakly stable if for every fluid model solution {D, T, Q} with Q(0)=0, Q(t)=0 for almost all $t \geq 0$.
\end{lemma}

Define Lyapunov function
\[
V(Q(t)) ~ = ~ \max_i \, Q_{i}(t)
\]
Note that in the definition of $V$ the maximum is taken over {\em all} ports, input and output.

{\bf Remarks}\\
\begin{itemize}
\item{
The Lyapunov function used by \cite{dimakis06} for the analysis of GMM policy also looks at the maximum queue length. The novelty of our proof is that we do not need to look at the individual queue lengths. Our Lyapunov function is based on port weights. Another difference is that while the analysis in \cite{dimakis06} depends on the fact that GMM is a maximal matching, our proof works for all LHPF policies which are not even required to be maximal, in general.
}
\item{
Our proof of stability is more subtle than the proof of stability for the MWM policy \cite{DaiPra}. Note that the maximum weight matching in the graph remains the maximum weight matching in the corresponding fluid model. However, the ports that are critical in a given interval of time $(t,t+\delta)$ in the fluid model may not be critical on a slot by slot basis in the actual system. Hence, for example, a critical port policy may not be able to schedule all the ports that are critical in the fluid model.
}
\end{itemize}
{\bf Proof Intuition}\\
{
Our proof is based on the observation that all the ports that are critical (heaviest) in the fluid limit, may not remain heaviest in the neighborhood of time $t$, but they continue to be above a certain threshold. We show that the optimal threshold must be below this threshold and hence all ports that are critical in the fluid limit are scheduled in {\em every} time-slot around $t$. We prove in Lemma~\ref{lem:fluid_critical} that a LHPF policy schedules all the ports that are critical in the corresponding fluid model and hence is throughput optimal. Note that the LHPF policy does not need to know which ports are critical in the fluid limit.
}

\begin{theorem}
\label{thm:fluid_decrease}
Any LHPF policy is throughput optimal.
\end{theorem}
{\em Proof:} Since V(Q(t)) is a non-negative function, to show that $V(t)=0$ for almost all $t \geq 0$, it is enough to show that, if  $t$ is a regular time and $V(t)>0$ then V(Q(t)) decreases at least at a given rate.
% We prove this in Theorem~\ref{thm:fluid_decrease}.

We prove that for all regular times $t$ such that $V(Q(t))>0$, for a system operating under any LHPF policy,
\[
\frac{d}{dt} V(Q(t)) \leq - \epsilon^*
\]

Fix time $t$ and let $\gamma = V(Q(t)) = \max_i \,
Q_i(t)$. Also, define
\[
\mC ~ = ~ \{i:Q_i(t) = \gamma \}
\]
to be the set of heaviest ports at $t$.  Also, let $\widetilde{\gamma}
= \max_{i\notin \mC} Q_i(t)$ be the heaviest of the remaining
ports. Since the number of ports is finite, $\widetilde{\gamma} <
\gamma$. Choose $\beta$ small enough so that {\em (a)}
$\widetilde{\gamma} < \gamma - 3\beta$, and {\em (b)} $\beta <
\frac{\gamma}{2N+1}$. Here $N = \max\{N_1,N_2\}$. Note that this
implies that
\begin{equation}\label{eq:beta}
\left ( \frac{N+1}{N} \right ) (\gamma-\beta) ~ > ~ \gamma + \beta
\end{equation}

Recall that $Q(t)$ is absolutely continuous. This means that there
exists a $\delta$ small enough, so that at all times $\tau\in
(t,t+\delta)$ the queues satisfy the following conditions
\begin{enumerate}
\item[(C1)] $Q_i(\tau) \in (\gamma-\frac{\beta}{2},\gamma+\frac{\beta}{2})$ for all $i\in \mC$
\item[(C2)] $Q_i(\tau) < \gamma-\frac{5\beta}{2}$ for all $i\notin \mC$
\end{enumerate}

Let $x_{k_l}$ be a positive subsequence for which the convergence to the fluid limit holds. Consider $l$ large enough so that $|\frac {q_{i}(x_{k_{l}}t)}{ x_{k_{l}}} - Q_{i}(t)| < \frac{\beta}{2}$. \\
Consider time slots $T := \{\lceil x_{k_l}t \rceil,  \lceil x_{k_l}t \rceil +1, \ldots \lfloor x_{k_l}(t+\delta) \rfloor \}$. The following lemma shows that all critical ports that are critical at the fixed time $t$ in the fluid limit will be scheduled at all time slots $n \in T$. The conditions (C1) and (C2) can be rewritten as follows for the original switching system.
\begin{enumerate}
\item[(C1*)] $q_i(n) \in [x_{k_l}(\gamma-\beta),x_{k_l}(\gamma+\beta)]$ for all $i\in \mC$
\item[(C2*)] $q_i(n) \leq x_{k_l}(\gamma-2\beta)$ for all $i\notin \mC$
\end{enumerate}

We state a lemma. We prove it immediately after the current proof.

\begin{lemma} \label{lem:fluid_critical}
For all times $n \in T$ , any LHPF policy will match all ports that are in $\mC$ at time $t$ in the fluid limit.
\end{lemma}

Now, assuming that a LHPF policy indeed schedules every port $i \in \mC$ at all times $n \in T$,
\begin{equation}
\sum_{M \in \mM} M_{i} (h_{M}(\lfloor x_{k_l}(t+\delta) \rfloor) - h_{M}(\lceil x_{k_l}t \rceil)) =  \lfloor x_{k_l}(t+\delta) \rfloor - \lceil x_{k_l}t \rceil \\
\end{equation}

Now by dividing both sides by $x_{k_{l}}$ and let $l \rightarrow \infty$, we obtain:
\begin{equation}
\label{help1}
\begin{split}
1 &\geq \frac { \sum_{M \in \mM} M_{i} ( h_{M}(x_{k_l}(t+\delta)) - h_{M} (x_{k_l}t) }{x_{k_{l}\delta}}\\
   &\geq \frac{\lfloor x_{k_l}(t+\delta) \rfloor - \lceil x_{k_l}t \rceil}{x_{k_{l}\delta} }\rightarrow 1
   \end{split}
\end{equation}

Hence for $\delta \rightarrow 0$,
\begin{equation}
\begin{split}
\sum_{M \in \mM}{M_{i}}\frac{d}{dt} H_{M}(t) = \lim_{\delta \rightarrow 0} \sum_{M \in \mM}{M_{i}} \frac{H_{M}(t+\delta) - H_{M}(t)}{\delta}\\
= \lim_{\delta \rightarrow 0} \lim_{l \rightarrow \infty} \frac{ \sum_{M \in \mM}{M_{i}} (h_{M}(x_{k_{l}}(t+\delta))-h_{M}(x_{k_{l}}t)}{x_{k_l}(\delta)} \\
         \rightarrow 1 \quad \quad \textrm{by Eq.~(\ref{help1})}\\
\end{split}
\end{equation}

So, by Eq.~(\ref{eq:der}) it follows that, $\forall i \in \mC$,\\
\begin{equation}
\displaystyle \frac{d Q_i(t)}{dt} = -(1-\lambda_i) \leq -\epsilon^*.
\end{equation}

Also, every port $i\notin \mC$ has weight strictly lower than every port in $\mC$, for the entire duration $(t,t+\delta)$. Thus it follows that
\[
\frac{d}{dt} V(Q(t)) ~ \leq ~ - \epsilon^*
\]
This proves the theorem. \hfill $\blacksquare$

{\em Proof of Lemma \ref{lem:fluid_critical}:}
%{\bf @@ this proof may need to be changed to go down to the level of slots}
%This proof is similar to the proof of Lemma \ref{lem:critical}.

Let $\mC_1 \subset \mC$ be the set of input ports in $\mC$, and $\mC_2 \subset \mC$ the set of
output ports.  We will first show that all ports in $\mC_1$ can be matched, by showing that Hall's condition (given in Lemma \ref{lem:halls}) holds for this set. By symmetry, all ports in $\mC_2$  can be matched and by Lemma~\ref{lem:merge} we conclude that all ports in $\mC$ can be matched.

Fix time $n \in T$, and for any subset $S\subset \mC_1$
let $\mN_\tau(S)$ be its neighborhood at time $n$. Suppose now
that $S$ fails Hall's condition, i.e. that $|S|\geq |\mN_{n}(S)| +
1$. Now, each $i\in S$ has $q_i(n) > x_{k_{l}}(\gamma-\beta)$, by condition
(C1). This means that
\[
\sum_{i\in S} q_i(n) ~ > ~ |S| x_{k_{l}}(\gamma-\beta) ~ \geq ~
\left ( |\mN_{n}(S)| +1 \right )x_{k_{l}}(\gamma-\beta)
\]
Now, each packet in $q_i(n)$, $i\in S$ is destined for one node
in $\mN_{n}(S)$, which means that
\[
\sum_{j\in \mN_{n}(S)} q_j(n) ~ \geq ~ \sum_{i\in S} q_i(n)
\]
(LHS and RHS may not be equal because there might be other input ports
with packets for ports in $\mN_{n}(S)$). This means that there exists
one node $j^*\in \mN_{n}(S)$ with queue
\begin{equation*}
\begin{split}
q_{j^*}(n) ~ \geq & ~ \frac{1}{|\mN_{n}(S)|} \sum_{j\in \mN_{n}(S)}
q_j(n) ~ \\
&> ~ \left ( \frac{|\mN_{n}(S)|+1}{|\mN_{n}(S)|}\right )
x_{k_{l}}(\gamma-\beta)
\end{split}
\end{equation*}

Further, $|\mN_{n}(S)| \leq N$, so we have that
\[
\frac{|\mN_{n}(S)|+1}{|\mN_{n}(S)|} ~ \geq ~ \frac{N+1}{N}
\]
which gives, from (\ref{eq:beta}),
\[
q_{j^*}(n) ~ > ~ \left ( \frac{N+1}{N} \right ) x_{k_{l}}(\gamma-\beta) >
x_{k_{l}}(\gamma + \beta)
\]
However, this means that $j^*$ violates the fact, implied by (C1*) and
(C2*), that $q_j(n) < x_{k_{l}}(\gamma + \beta)$ for all ports $j$.
This is a contradiction, and thus it has to be that Hall's condition is
satisfied at $n$.

Thus, there exists a matching that matches all input ports in
$\mC_1$. Similarly, it can be shown there exists a matching that
matches all output ports in $\mC_2$. By Lemma \ref{lem:merge}, this
means there exists a matching that matches all ports in $\mC$. From
conditions (C1*) and (C2*), it follows that this matching matches all
ports with weight greater than $\gamma-\beta$. So its threshold is
$\gamma-\beta$, or lower. Now, by definition of LHPF, this means that
the threshold of any LHPF matching cannot be greater than
$\gamma-\beta$. This means the LHPF matching schedules all ports with
weights above $\gamma-\beta$, i.e. all ports in $\mC$. \hfill $\blacksquare$

\begin{figure}[t]
\centering
\epsfig{file = 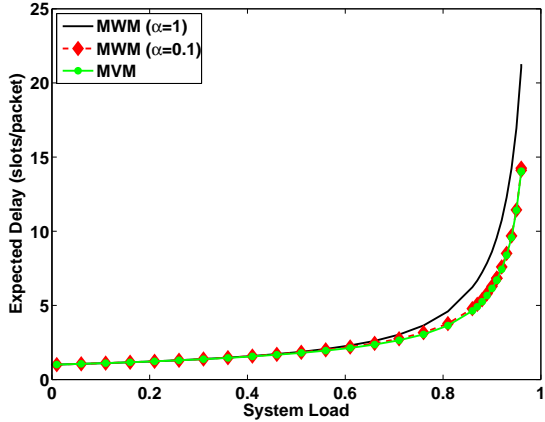, width = 3.3 in}
\caption{\footnotesize Delay of a 8 $\times$ 8 switch under symmetric Bernoulli Traffic}
\label{fig:ber}
\end{figure}

\section{Simulations}
In this section we compare the delay performance of the MVM algorithm with MWM-$\alpha$ algorithms.
MVM lies in the class of LHPF policies. We implemented a packet level simulator in Java.  The simulations are run long enough so that the half-width of the 99\% confidence interval is within 1\% of the mean.

We simulate a 8 $\times$ 8 switch with symmetric loading on each edge. We simulate two types of arrival processes, Bernoulli and a more bursty arrival process. Each arrival stream injects packets independently in the system. Clearly, these processes satisfy strong law of large numbers and the switch is guaranteed to be stable. The model of the bursty arrival process is described below.

{\bf Bursty Arrival Processes:} The arrival stream is a series of active and idle periods. During the active periods, the source injects one packet into the queue in every time slot. The length of the active periods (denoted by random variable $a$) are distributed according the Zipf law with power exponent 1.25 and support [1,2,3,\ldots,100]. Heavy tailed distributions like Zipf, have been found to model the Internet traffic \cite{IMC04}. During the active period the source generates one packet every time-slot. The idle periods are geometrically distributed with mean $p$. The mean arrival rate of a source can be controlled by changing the value of $p$.

The results for Bernoulli traffic in Fig.~\ref{fig:ber} show that the delay of MVM policy is smaller than that of the MWM policy. MWM-$\alpha$ policies have been studied in the literature \cite{Mc01, shah06} and have been reported to incur smaller delay as the value of $\alpha$ goes to zero. Our simulations confirm this observation and also show that the delay performance of the MVM policy is no worse than the MWM-$\alpha$ policies even for small values of $\alpha$.

\begin{figure}[t]
\centering
\epsfig{file = 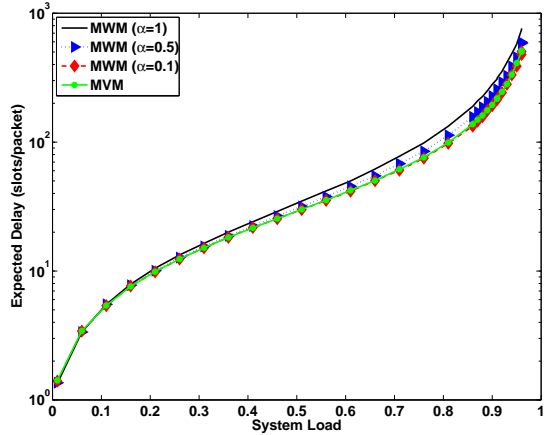, width  = 3.3 in}
\caption{\footnotesize Delay of a 8 $\times$ 8 switch under symmetric Bursty Traffic}
\label{fig:zipf}
\end{figure}

Fig.~\ref{fig:zipf} shows the delay for the bursty arrival process described above. The delay is significantly higher for the more bursty arrival process as compared to Bernoulli traffic. It seems that although the MVM and MWM-0+ policies have different tie-breaking rule, their delay performance is actually quite similar.

\section{Discussion}

This paper proposes a new class of online policies called LHPF policies for scheduling in input-buffered crossbar switches. LHPF policies are both throughput optimal for a system with arrivals, and clearance-time optimal for a system without arrivals. To our knowledge, this is the first class of online policies that achieves both objectives. We also provide necessary and sufficient conditions for any policy to be clearance time optimal, and show that popular existing policies (like MWM, MSM and Greedy/GMM) can have clearance-time as large as twice the optimal. A particular policy in our class, MVM, has worst-case complexity similar to MSM (which is not known to be throughput-optimal), and empirical delay performance better than MWM. 

As noted in \cite{Mc98}, the MVM policy combines the benefit of a maximum size algorithm with those of a maximum weight algorithm, while lending itself to simple implementation in hardware. In MVM, each weight is a function of queue lengths (sum of all edges that touch a node) and hence has a advantage of both the maximum size matchings with high instantaneous throughput while guaranteeing high throughput, even when the arrival traffic is non-uniform. The LHPF class of policies do not care about the weight of edges as far as the required set of nodes above the optimal threshold have been scheduled. This reduces the computational overhead for the scheduler while maintaining the throughput guarantee.

%The delay analysis of the throughput optimal schedulers in the fluid limits is still limited to MWM type algorithms \cite{Sasha04, shah06}. The Lyapunov function designed here is much simpler, and opens avenues for the development of fluid analysis of such systems.

Philosophically, this paper departs from the prevalent edge-based approach to scheduling, as exemplified by MWM (schedule the heaviest queues), MSM (biggest number of queues) or Greedy. Instead we concentrate on the most congested ports. It would be interesting to see if the results of this paper generalize to other interference models (e.g. those that arise in wireless networks). In particular, ports in switches represent the scheduling constraints (at most one edge per port can be scheduled). More generally, we might concentrate on the most-congested {\em constraints}, like e.g. cliques in the conflict-graph setting. For such a setting the Lyapunov function may be the heaviest constrained set. 

Our Lyapunov function is evocative of the one used by \cite{dimakis06} for the analysis of Greedy (GMM). However, we emphasize that ours is a function of the total queues at ports, while theirs is of every single queue. Besides, the Lyapunov function of \cite{dimakis06}, other popular Lyapunov functions are also all based on individual queue lengths: sum of squares of queue lengths (for MWM) etc.

In the fluid limit, \cite{dimakis06} can guarantee that among the set of critical queues, a maximal set of queues can be scheduled at every time slot. The GMM policy is throughput optimal only when the underlying graph satisfies a local pooling condition. Note that the GMM policy is not throughput optimal for switches. In contrast, using the node based formulation, we have been able to prove that LHPF policies are throughput optimal because they guarantee that every port that is critical in the fluid limit can be scheduled at every time slot. This is because of the special structure (bipartite) graph of a switch. More generally, it has been shown that the GMM policy achieves at least a portion of the capacity region which is given by the local-pooling factor \cite{Joo07, Joo08, Bree08}. It would be very interesting to see if this approach would lead to the development of simpler policies with throughput guarantees for more general class of networks; especially since MWM matching problem although throughput optimal, has exponential complexity in the general setting.

%which should have the following theme: "it is worth investigating if
%node-weighted matching ideas will become as widely applicable as
%max-weight scheduling". speak first about the generality of max-weight
%policies, how variants like pick-and-compare etc. have ben developed, and
%how they are so useful. (you already have someof the material for this, i
%would just avoid excessive referencing - thatis for the related work
%section). some candidate sentences (use at own discretion):

%More generally, the GMM policy achieves a certain fraction of the 

%Switch scheduling has been studied extensively both because of its relevance to high speed switching and its pedagogical example as a network complex enough to inspire interesting research yet simple enough for an extensive network theory to be developed. Minimum clearance time policies have been applied to stabilize networks in \cite{Neely02a, Gamarnik}.
%In addition to IQ switches, there are many other examples of
%scheduling in various problems arising in networks, that have
%throughput optimal policies as ¡°maximum weight matching¡±
%type algorithms. For example, Max-pressure policy for network
%of queues [7], scheduling in Radio hop network [4],
%Generalized MWM [8] and scheduling in network of switches
%[9]. In summary, MWM type algorithms are the heart of good
%scheduling solutions in many network applications.
%The MWM algorithm for example is throughput optimal, not only for the switch network with cross bar constraints, but more general network like wireless networks with interference constraints.

\bibliographystyle{acm}
\bibliography{Gagan}

\begin{thebibliography}{10}

\bibitem{Altman}
{\sc Altman, E., Liu, Z., and Righter, R.}
\newblock Scheduling of an input-queued switch to achieve maximal throughput.
\newblock {\em Probability in the Engineering and Informational Sciences 14\/}
  (2000), 327--334.

\bibitem{Bree08}
{\sc Brzezinski, A., Zussman, G., and Modiano, E.}
\newblock Enabling distributed throughput maximization in wireless mesh
  networks: A partitioning approach.
\newblock In {\em ACM Mobicom\/} (New York, NY, USA, 2006), ACM Press,
  pp.~26--37.

\bibitem{Chang}
{\sc Chang, C.-S., Chen, W.-J., and Huang, H.-Y.}
\newblock On service guarantees for input buffered crossbar switches: A
  capacity decomposition approach by birkhoff and von neumann.
\newblock In {\em IEEE Infocom\/} (2000).

\bibitem{DaiPra}
{\sc Dai, J., and Prabhakar, B.}
\newblock The throughput of data switches with and without speedup.
\newblock In {\em IEEE Infocom\/} (March 2000).

\bibitem{Dai}
{\sc Dai, J.~G.}
\newblock On positive harris recurrence of multiclass queueing networks: a
  unified approach via fluid limit models.
\newblock {\em Annals of Applied Probability 5\/} (1995), 49--77.

\bibitem{dimakis06}
{\sc Dimakis, A., and Walrand, J.}
\newblock Sufficient conditions for stability of longest-queue-first
  scheduling: Second-order properties using fluid limits.
\newblock {\em Advances in Applied Probability 38}, 2 (2006), 505–--521.

\bibitem{IMC04}
{\sc Feldmann, A., Kammenhuber, N., Maennel, O., Maggs, B., Prisco, R.~D., and
  Sundaram, R.}
\newblock A methodology for estimating interdomain web traffic demand.
\newblock In {\em IMC\/} (2004).

\bibitem{NOW}
{\sc Georgiadis, L., Neely, M.~J., and Tassiulas, L.}
\newblock {\em Resource Allocation and Cross-Layer Control in Wireless
  Networks, Foundations and Trends in Networking}, vol.~1.
\newblock Now Publishers, 2006.

\bibitem{Kuhn}
{\sc Hoffman, A., and Kuhn, H.}
\newblock Systems of distinct representatives and linear programming.
\newblock {\em The American Mathematical Monthly 63\/} (1956).

\bibitem{Mc02}
{\sc Iyer, S., and McKeown, N.}
\newblock Maximum size matching and input queued switches.
\newblock In {\em Proceedings of the 40th Annual Allerton Conference on
  Communication, Control and Computing\/} (2002).

\bibitem{Joo07}
{\sc Joo, C., Lin, X., and Shroff, N.~B.}
\newblock Performance limits of greedy maximal matching in multi-hop wireless
  networks.
\newblock In {\em IEEE CDC\/} (2007).

\bibitem{Joo08}
{\sc Joo, C., Lin, X., and Shroff, N.~B.}
\newblock Understanding the capacity region of greedy maximal scheduling
  algorithm in multi-hop wireless networks.
\newblock In {\em IEEE INFOCOM\/} (2008).

\bibitem{karp}
{\sc Karp, R., and Hopcroft, J.}
\newblock An n/sup 5/2/ algorithm for maximum matchings in bipartite graphs.
\newblock {\em SIAM Journal on Computing 2\/} (1973).

\bibitem{Mc01}
{\sc Keslassy, I., and McKeown, N.}
\newblock Analysis of scheduling algorithms that provide 100\% throughput in
  input-queued switches.
\newblock In {\em Proceedings of the 39th Annual Allerton Conference on
  Communication, Control, and Computing\/} (October 2001).

\bibitem{Mc03}
{\sc Keslassy, I., Zhang-Shen, R., and McKeown, N.}
\newblock Maximum size matching is unstable for any packet switch.
\newblock {\em IEEE Communications Letters 7\/} (October 2003).

\bibitem{leonardi01}
{\sc Leonardi, E., Mellia, M., Neri, F., and M., A.~M.}
\newblock On the stability of input-queued switches with speed-up.
\newblock {\em IEEE/ACM Transactions on Networking 9}, 1 (Feburary 2001).

\bibitem{Hall}
{\sc M.~Hall, J.}
\newblock Distinct representatives of subsets.
\newblock {\em Bulletin of the American Mathematical Society 54\/} (1948).

\bibitem{Mc96}
{\sc McKeown, N., Anantharam, V., and Walrand, J.}
\newblock Achieving 100\% throughput in an input-queued switch.
\newblock In {\em IEEE Infocom\/} (March 1996).

\bibitem{Mc98}
{\sc Mekkittikul, A., and McKeown, N.}
\newblock Practical scheduling algorithm to achieve 100\% throughput in
  input-queued switches.
\newblock In {\em IEEE Infocom\/} (April 1998).

\bibitem{Neely07}
{\sc Neely, M.~J., Modiano, E., and Cheng, Y.-S.}
\newblock Logarithmic delay for n x n packet switches under the crossbar
  constraint.
\newblock {\em IEEE Transactions on Networking 15\/} (2007).

\bibitem{Neely02c}
{\sc Neely, M.~J., Modiano, E., and Rohrs, C.~E.}
\newblock Tradeoffs in delay guarantees and computation complexity for n¡¿n
  packet switches.
\newblock In {\em Proc. of Conf. on Information Sciences and Systems (CISS)\/}
  (2002).

\bibitem{Gamarnik}
{\sc Neely, M.~J., Sun, J., and Modiano, E.}
\newblock Stability of adaptive and non-adaptive packet routing policies in
  adversarial queueing networks.
\newblock In {\em Proc. of 31st ACM Symposium on the Theory of Computing\/}
  (1999).

\bibitem{Neely02a}
{\sc Neely, M.~J., Sun, J., and Modiano, E.}
\newblock Delay and complexity tradeoffs for routing and power allocation in a
  wireless network.
\newblock In {\em Proc. of Allerton Conf. on Communication, Control, and
  Computing\/} (2002).

\bibitem{Perfect}
{\sc Perfect, H., and Pym, J.}
\newblock An extension of banach's mapping theorem, with applications to
  problems concerning common representatives.
\newblock {\em Proceedings of the Cambridge Philosophical Society (Mathematical
  and Physical Sciences) 62\/} (1966).

\bibitem{shah03}
{\sc Shah, D.}
\newblock Maximal matching scheduling is good enough.
\newblock In {\em IEEE Globecom\/} (2003).

\bibitem{shah02}
{\sc Shah, D., Giaccone, P., and Prabhakar, B.}
\newblock An efficient randomized algorithm for input-queued switch scheduling.
\newblock {\em IEEE Micro 22\/} (January-February 2002).

\bibitem{shah06}
{\sc Shah, D., and Wischik, D.}
\newblock Optimal scheduling algorithms for input-queued switches.
\newblock In {\em INFOCOM\/} (2006).

\bibitem{Sasha04}
{\sc Stolyar, A.~L.}
\newblock Maxweight scheduling in a generalized switch: State space collapse
  and workload minimization in heavy traffic.
\newblock {\em Annals of Applied Probability Vol.14}, No.1 (2004), pp.1--53.

\bibitem{Tass}
{\sc Tassiulas, L.}
\newblock Scheduling and performance limits of networks with constantly
  changing topology.
\newblock {\em IEEE Trans. Inform. Theory 43\/} (1997), 1067--1073.

\bibitem{tassiulas92}
{\sc Tassiulas, L., and Ephremides, A.}
\newblock Stability properties of constrained queueing systems and scheduling
  policies for maximum throughput in multihop radio networks.
\newblock {\em IEEE Trans. Aut. Contr.37 37}, 12 (1992), 1936--1948.

\bibitem{weller}
{\sc Weller, T., and Hajek, B.}
\newblock Scheduling nonuniform traffic in a packet-switching system with small
  propagation delay.
\newblock {\em IEEE/ACM Trans. Netw. 5}, 6 (1997), 813--823.

\end{thebibliography}
\balancecolumns
\end{document}